\documentstyle[12pt]{article}

\newlength{\dinwidth}
\newlength{\dinmargin}
\begin{document}

\def\bold#1{\setbox0=\hbox{$#1$}%
     \kern-.025em\copy0\kern-\wd0
     \kern.05em\copy0\kern-\wd0
     \kern-.025em\raise.0433em\box0 }
\def\slash#1{\setbox0=\hbox{$#1$}#1\hskip-\wd0\dimen0=5pt\advance
       \dimen0 by-\ht0\advance\dimen0 by\dp0\lower0.5\dimen0\hbox
         to\wd0{\hss\sl/\/\hss}}
\def\lq{\left [}
\def\rq{\right ]}
\def\LL{{\cal L}}
\def\VV{{\cal V}}
\def\AA{{\cal A}}
\def\BB{{\cal B}}
\def\MM{{\cal M}}

\newcommand{\be}{\begin{equation}}
\newcommand{\ee}{\end{equation}}
\newcommand{\bea}{\begin{eqnarray}}
\newcommand{\eea}{\end{eqnarray}}
\newcommand{\nn}{\nonumber}
\newcommand{\dd}{\displaystyle}
\newcommand{\bra}[1]{\left\langle #1 \right|}
\newcommand{\ket}[1]{\left| #1 \right\rangle}
\newcommand{\spur}[1]{\not\! #1 \,}
\thispagestyle{empty}
\vspace*{1cm}
\rightline{BARI-TH/95-203}
\rightline{June 1995}
\vspace*{2cm}
\begin{center}
  \begin{Large}
  \begin{bf} 
Leptonic constant from $B$ meson  radiative decay\\
  \end{bf}
  \end{Large}
  \vspace{8mm}
  \begin{large}
P. Colangelo $^{a,}$ \footnote{E-mail address:
COLANGELO@BARI.INFN.IT},  F. De Fazio $^{a,b}$, G. Nardulli $^{a,b}$\\
  \end{large}
  \vspace{6mm}
$^{a}$ Istituto Nazionale di Fisica Nucleare, Sezione di Bari, Italy\\
  \vspace{2mm}
$^{b}$ Dipartimento di Fisica, Universit\'a 
di Bari, Italy \\

\end{center}
\begin{quotation}
\vspace*{1.5cm}
\begin{center}
  \begin{bf}
  Abstract\\
  \end{bf}
\end{center}

We propose a method to determine the leptonic decay constant $f_{B^*}$ 
in the infinite heavy quark mass limit  from the analysis of the 
radiative decay mode 
$B^- \to \ell^- {\bar \nu}_\ell \gamma$. 
The method relies on HQET symmetries and on experimental data from $D^{*0} \to 
D^0 \gamma$.

\vspace*{0.5cm}
\end{quotation}

\newpage
\baselineskip=18pt
\setcounter{page}{1}

The measurement of $f_B$ and  $f_{B^*}$, the leptonic decay constants 
of the $B$ and $B^*$ mesons, defined by the matrix elements:
\begin{equation}
<0|{\bar b}\gamma_\mu \gamma_5 q|B(p)>=ip_\mu f_B \hskip 3 pt  \label{fb}
\end{equation}
\noindent 
\begin{equation}
<0|{\bar b}\gamma_\mu q|B^*(p, \epsilon)>=m_{B^*} f_{B^*} \epsilon_\mu 
\hskip 3 pt , \label{fbstar} \end{equation}
\noindent
represents one of the main goals of the current and future experimental 
investigations in the heavy quark physics. The reason can be found in the prime
role played by $f_B$ in the hadronic systems containing one heavy quark.
To give an example, $f_B^2$ appears in the formula relating the mass difference 
between the $B^0$-meson mass eigenstates to $|V_{td}|^2$ in the box diagram 
computation of the $B^0-{\bar B}^0$ mixing; therefore, the size of the 
unitarity 
triangle and the analysis of possible CP violations in the $B$ systems 
crucially depend on the value of this hadronic parameter. \par
As a second example of the relevance of the leptonic $B$ constant,
we can consider the 
Heavy Quark Effective Theory applied to the physical world of the 
heavy hadrons. 
In the limit $m_b \to \infty$, $f_B$ 
and $f_{B^*}$ scale according to the relation
\begin{equation}
f_B=f_{B^*}={\hat{F} \over \sqrt{m_B}} \hskip 3 pt . \label{hat}
\end{equation}
\noindent 
The parameter $\hat{F}$, independent (modulo logarithmic corrections) of 
the heavy quark mass, represents a low-energy parameter related to the 
non-perturbative dynamics of light quark and gluon degrees of freedom, and 
plays  a role analogous to $f_\pi$ in chiral 
theories for light hadrons.

On the theoretical side, much effort has been devoted to the calculation of 
$f_B={\hat{F} \over \sqrt{m_B}}$ by non-perturbative methods such as
lattice QCD \cite{reticolo} and QCD sum 
rules \cite{QCDsr}; for example, a QCD sum rules analysis in the infinite 
heavy quark mass limit provides us with the value:
\begin{equation}
\hat{F}= 0.25 - 0.45 \; GeV^{3 \over 2}\hskip 3 pt . \label{neubert} 
\end{equation}
\noindent
(depending on the inclusion of $\alpha_s$ corrections).

On the experimental side, the most natural process to measure $f_B$ would be 
the purely leptonic decay mode $B^- \to \ell^- {\bar \nu_\ell}$, whose 
decay width is given by:
\begin{equation}
\Gamma(B^- \to \ell^- {\bar \nu_\ell})={G_F^2 \over 8 \pi} |V_{ub}|^2 f_B^2 
\Bigg( {m_\ell \over m_B} \Bigg)^2 m_B^3 \Bigg( 1-{m_\ell^2 \over m_B^2} 
\Bigg)^2 \hskip 3 pt . \label{0} 
\end{equation}
\noindent If one determines $|V_{ub}|$ from other processes, for 
example from the end-point spectrum of the charged lepton in the $B$ meson 
inclusive semileptonic decay \cite{shifman,hon}, one
can obtain $f_B$ by
this equation. The difficulty is 
represented by the helicity suppression displayed in Eq. 
(\ref{0}). Since  the 
lepton pair must be a spin 0 state and the antineutrino has a 
right-handed helicity, also 
the charged lepton is forced to be right-handed. The effect is the 
suppression factor   $\big( { m_\ell \over m_B}  \big)^2$
that makes the purely leptonic decay mode hardly 
accessible in the electron and in the muon case. As a matter 
of fact, using the value for the $B^-$ lifetime
$\tau_{B^-}=1.646 \pm 0.063 \; ps$ \cite{hon},
the expected rates are: 
\bea
 {\cal B}(B^- \to e^- {\bar \nu_e}) & \simeq& \; 6.6 \; 
\big[{V_{ub} \over 0.003} \big]^2
\big[{f_B \over 200 \; MeV} \big]^2 \; 10^{-12}  \\
 {\cal B} (B^- \to \mu^- {\bar \nu_\mu}) & \simeq & \; 2.8 \; 
\big[{V_{ub} \over 0.003} \big]^2
\big[{f_B \over 200 \; MeV} \big]^2 \; 10^{-7} \; ,
\eea
to be compared with the experimental upper bound put by CLEO
\cite{CLEOIIa}: 
\be {\cal B} (B^- \to e^- {\bar \nu_e})<1.5 \; 10^{-5}\ee 
\be {\cal B} (B^- \to \mu^- {\bar \nu_\mu})<2.1 \; 10^{-5}\ee
(at $90 \; \%$ confidence level). \par
As for the channel $B \to \tau \nu_\tau$, the helicity suppression 
is absent
and  the expected rate is larger: 
\be
{\cal B} (B^- \to \tau^- {\bar \nu_\tau}) \simeq \;
6.8 \; \big[{V_{ub} \over 0.003} \big]^2
\big[{f_B \over 200 \; MeV} \big]^2 \; 10^{-5} \; .
\ee
However, the 
$\tau$ identification puts a hard experimental challenge.
The upper limits found  by CLEO \cite{CLEOIIa}  
and ALEPH \cite{aleph}
Collaborations read:
\bea
{\cal B}(B^- \to \tau^- {\bar \nu_\tau}) & < &  2.3 \; 10^{-3} \\
{\cal B}(B^- \to \tau^- {\bar \nu_\tau}) & < & 1.8 \; 10^{-3}  
\eea
(at $90 \; \%$ confidence level), respectively.

For these reasons it is worthwhile to search for other paths, and 
analyze  other possible decay modes that are sensitive to the value of 
$\hat{F}$. For example, one could use flavour symmetry and
consider a measurement of
$f_{B^*}$ from the spectrum of the semileptonic  
decay $B \to \pi  \ell \nu_\ell$ near  zero recoil 
\cite{ligeti,noi1} compared to the spectrum of 
$D \to \pi  \ell \nu_\ell$ in the same kinematics.
In this case, however, one has to face a 
strong phase-space suppression.

Another possibility, suggested in refs.\cite{atwood} and \cite{burdman}, is 
provided by the radiative leptonic decay channel
\begin{equation}
B^- \to \mu^- {\bar \nu_{\mu}} \gamma \hskip 3 pt  \label{channel} 
\end{equation}
\noindent which does not suffer of the helicity suppression because of the 
presence of the photon in the final state.
In this case there are several uncertainties  related to the 
hadronic parameters appearing in the matrix element governing the decay 
mode (\ref{channel}). Within such uncertainties, the branching ratio of
the decay (\ref{channel}) has 
been estimated in the range $10^{-7}-10^{-6}$ (in the case of light leptons), 
thus making the channel promising for a future $B$-factory. We shall now study 
how this decay mode can be used to measure $f_{B^*}$.

In order to analyze the dependence of the amplitude for the 
process (\ref{channel}) on $f_{B^*}$, let us consider the diagrams
which describe it; they can be divided into two classes.
The first class consists of structure dependent (SD) 
diagrams such as those of Figs. 1; the second class contains 
bremsstrahalung diagrams where the photon is emitted from 
the $B^-$ or from the charged lepton leg.

The bremsstrahalung amplitude  is given by:
\begin{equation}
{\cal M}_{B}=i f_B e {G_F \over \sqrt{2}}V_{ub}m_{\mu} \big[\big( F(k^2) 
{\epsilon \cdot p \over p \cdot k} -{\epsilon \cdot p_\ell \over p_{\it l} 
\cdot k}\big) {\bar {\mu}}(1-\gamma_5)\nu -
{1 \over 2 p_{\it l} \cdot k} {\bar {\mu}} {\spur \epsilon} \spur k 
(1-\gamma_5)\nu \big] \hskip 3 pt \label{1} \end{equation} 
\noindent where $p$, $p_{\it l}$, $k$ are the momenta of $B^-$, $\mu$ and 
$\gamma$, respectively ,
$\epsilon $ is the photon polarization vector, and $F(k^2)$ is the $B^-$ 
electromagnetic form factor. This
contribution vanishes in the limit $m_\mu \to 0$ and we shall make this 
approximation, so that the relevant 
diagrams governing (\ref{channel}) 
are the SD diagrams. 
We shall suppose, as in \cite{burdman}, 
that in these polar diagrams the intermediate state can be a 
$J^P=1^- (B^*)$ and a positive parity $B^{**}$ meson.
The amplitude with intermediate $P(=B^*, B^{**})$ state
can be written as follows:
\begin{equation}
{\cal M}_{SD}^{(P)}= {G_F \over \sqrt{2}} V_{ub} 
{\cal A}(B \to P \gamma) 
{i \over (p-k)^2-m^2_P} <0|{\bar u} \gamma^{\mu} (1-\gamma_5)b|P>
{\it l}_{\mu} \hskip 3 pt , \label{2} 
\end {equation}
\noindent where ${\it l}_{\mu}={\bar \ell}(p_{\it l}) \gamma_{\mu}(1-\gamma_5) 
\nu(p_{\nu})$ is the lepton current, ${\cal A}(B \to P \gamma)$ is the 
amplitude of the process $B \to P \gamma$, and $P$ indicates the pole. 
From Eq.(\ref{2}) (with $P=B^*$) 
it is clear that, for light leptons in the final 
state, the radiative $B$ decay can give 
access to the decay constant $f_{B^*}$ provided that \\
1) one has enough information on the amplitude  ${\cal A}(B^* \to B \gamma)$, \\
2) the remaining part of the SD contribution ${\cal M}_{SD}^{(B^{**})}$ is 
small compared to ${\cal M}_{SD}^{(B^{*})}$.

Let us begin by discussing the first point and let us consider the 
contribution  of the $B^*$ pole to (\ref{2}). In computing the on-shell 
amplitude ${\cal A}(B^* \to B \gamma)$ one has to take into account 
the coupling of the electromagnetic 
current to the heavy quark and to the 
light quark, i.e. the terms arising from the decomposition 
$J_\mu^{em}=   e_b {\bar b} \gamma_\mu b + e_q {\bar q} \gamma_\mu q$ 
($e_b, e_q=$ quark charges).

The $b-$quark contribution is described 
(in the limit $m_b \to \infty$) by the amplitude:
\begin{equation}
\epsilon^{* \mu}
<B(v^\prime)|e_b {\bar b} \gamma_\mu b |B^*(v, \eta)>=i \; e_b 
\;  \xi(v \cdot v^\prime) \sqrt{m_B m_{B^*}}\; \epsilon_{\mu \nu \alpha \beta} 
\;\epsilon^{* \mu}
\; \eta^\nu v^\alpha v^{\prime \beta} \hskip 3 pt , \label{2.1} 
\end{equation}
\noindent where 
$v$ and $v^\prime$ are $B^*$ and $B$ four-velocities, respectively,
 and $\xi(v \cdot v^\prime)$ is the universal Isgur-Wise form factor (IW) 
\cite{isgur}. For on-shell $B$ and $B^*$, 
since $v \cdot v^\prime= { m_B^2 + m_{B^*}^2 \over 2 m_B m_{B^*}} \simeq 1$, 
we can use the normalization of the IW 
function: $\xi(1)=1 $.

The second contribution: 
$\epsilon^{* \mu} \;  <B|e_q \; {\bar q} \gamma_\mu q|B^*>$ 
represents the coupling of the
electromagnetic current to the 
light quark $q$ ($q=u$ for $B^-$ decay); 
this contribution dominates in the $m_b \to \infty$ limit, and is more 
uncertain since it cannot be estimated
within HQET because it involves light degrees of freedom.
On the experimental side we have no data on the width
$B^* \to B \gamma$ at the moment, and it is unlikely it will be 
measured in the near future. On the other hand, the experimental 
$D^{*0,-} \to D^{0,-} \gamma$ branching ratios  are known 
(even though the full $D^{*0,-}$ width
has not been measured yet) and we may presume that in future
we will get information on the $D^*$ partial radiative width. 

Our proposal is to use these pieces of information
to obtain $B^* \to B \gamma$.

The $D^{*0}$ radiative width is given by:
\be
\Gamma(D^{*0} \to D^0 \gamma) = { q_\gamma^3 \over 12 
\pi}\frac{ m_{D^{*0}}}{m_{D^0}} g^2_{D^* D \gamma} \label{gamma}
\ee 
where $q_\gamma$ is the photon momentum in
the $D^{*}$ rest frame and 
\be
g_{D^* D \gamma}=e \Big[\frac{e_c}{m_{D^*}} +
\frac{e_q}{\Lambda_q} \Big]  \label{gdg}
\ee
\noindent
($e_c=2/3, \, e_q=e_u=2/3$).
A measurement of  $\Gamma(D^{*0} \to D^0 \gamma)$ would provide
a determination of the mass constant $\Lambda_q$ that parametrizes
the matrix element $<D|{ \bar q} \gamma_\mu q |D^*>$. On the other
hand
\begin{eqnarray}
{\cal A}(B^*(v,\eta) \to B(v^{\prime}) \gamma(q, \epsilon))= 
&i& \; e \; \left[{e_b \over m_{B^*}}  + {e_q \over \Lambda_q} 
\right]  \times \nonumber 
\\
&& m_{B^*} \sqrt{m_B m_{B^*}} \epsilon_{\mu \nu \alpha \beta} 
\epsilon^{* \mu} \eta^{\nu}
v^{\alpha} v^{\prime \beta}  \hskip 3 pt , \label { 3} 
\end{eqnarray}
\noindent 
and therefore, by the the knowledge
of $\Lambda_q$, one would obtain the matrix element needed to
compute Eq.(\ref{2}) (with $P=B^*$). In other words with the
approach we have just described, we might extract ${\hat F}$
by the measurement of {\it both}
$\Gamma(D^{*0} \to D^* \gamma)$ and $\BB(B^- \to \mu^- {\bar \nu}_\mu \gamma)$; 
we shall discuss below the sensitivity of this approach.

In order to assess the reliability of this method we shall also discuss
as a consistency test
based on information in part already
available, i.e. data on $B \to \pi \ell \nu$ \cite{CLEOII}, $D \to K \ell 
\nu$ \cite{PDG}
and on the $BR'$s $\BB(D^{*} \to D \pi)$ 
and $\BB(D^{*} \to D \gamma)$ \cite{PDG}. Let us
consider the partial width
\be
\Gamma(D^{*0} \to D^0 \pi^0) = \Big( {g_{D^* D \pi} \over \sqrt 2} \Big)^2 
{ q_\pi^3 \over 24 \pi m^2_{D^*}} \label{pione}
\ee
where $q_\pi$ is the pion momentum in the $D^*$ rest frame and
  $g_{D^* D \pi}$ is the strong   ${D^* D \pi}$  coupling constant, which,
in the heavy quark mass limit, is given by
\be
g_{D^{*} D \pi} = g_{D^{*+} D^0 \pi^+} = \frac{2 \, g}{f_\pi} 
\sqrt{m_D m_{D^*}} \; .\ee
The ratio of the partial widths (\ref{pione}) and (\ref{gamma}) is
experimentally known \cite{PDG}
\be
R=\frac{ \BB(D^{*0} \to D^0 \pi^0) }{\BB(D^{*0} \to D^0 \gamma)}
= 1.75 \pm 0.21 \; ,
\ee
from which the ratio of the $D^*$ decay constants can be obtained:
\be
\frac{g}{g_{D^{*} D \gamma}} = 0.808 \; \sqrt{R} \; GeV \; . \label{ratio}
\ee
In order to extract $g_{D^{*} D \gamma} $ from Eq.(\ref{ratio})
one needs the strong coupling constant $g$; several
determinations of $g$ have appeared in the literature, based on
QCD sum rules \cite{noi1,grozin,belyaev} or potential models \cite{fdf}; they
indicate a value in the range $0.2 \, - \, 0.4$.
We shall avoid, however, to rely on these
estimates and we shall try to minimize the theoretical bias 
using experimental data of the decay
 $D \to K \ell \nu$ ($\BB(D^+ \to {\bar K}^0 e^+ \nu_e)=
6.6 \pm 0.9 \; 10^{-2}$ 
\cite{PDG}) and the recently observed decays
\cite{CLEOII}
\bea
\BB(B \to \pi \ell \nu) & =& 1.70 \pm 0.50  \; 10^{-4} ~~~~~~~~~(BSW) 
\label{bsw} \\
\BB(B \to \pi \ell \nu) & =& 1.19 \pm 0.65  \; 10^{-4} ~~~~~~~~~(ISGW) \; .
\label{isgw}
\eea
\noindent
The two determinations refer to the model used
in the Montecarlo code to compute the efficiencies: the Bauer et al.
model \cite{bsw} or the Isgur et al. model \cite{isgw}.

For these decays,  assuming a simple pole formula for the form factors
$f_+^{B \to \pi}(q^2)$ and 
$f_+^{D \to K}(q^2)$, 
which is what is experimentally found for $D^+ \to {\bar K}^0 \ell^+ \nu_\ell$ 
and generally accepted on theoretical grounds \cite{ball, santorelli, mende, 
casal} for $B \to \pi \ell \nu$, one gets the following results.
The semileptonic $B\to \pi \ell \nu$ partial width is given by
\be
\Gamma(B \to \pi \ell \nu ) = {\hat F}^2 g^2 |V_{ub}|^2
\frac{G_F^2}{192 \pi^3 f^2_{\pi}  m^4_B} J
\label{gammab}
\ee
with
\be
J= \int_0^{q^2_{max}} dq^2 \;  \frac{\lambda^{3/2}(m_B^2, m_\pi^2, q^2)}
{(1-q^2/m^2_{B^*})^{2}} \; . \label{j}
\ee
In (\ref{j}) $\lambda$ is the triangular function. For $D \to K \ell 
\nu_\ell$ one has to change $m_B \to m_D$, $V_{ub} \to V_{cs}$,
 $m_{B^*} \to m_{D^*_s}$, $m_\pi \to m_K$. From Eq.(\ref{j})
one  gets:
\be
{\hat F} g  = {2.95 \;  10^{-2}\over |V_{ub}|} \; \Bigg[
\BB ({\bar B^0} \to \pi^- \ell \nu) \Bigg]^{1/2} GeV^{3 \over 2}=  
(1.07 \; - \; 1.28 ) \; 10^{-1} \; {1\over |{V_{ub}\over 0.003}|}\; GeV^{3/2} , 
\label{res}
\ee
\be
{\hat F} g  = { 0.45 \over |V_{cs}|} \; \Bigg[
\BB ({\bar D^+} \to K^0 e^+ \nu_e) \Bigg]^{1/2} GeV^{3 \over 2}=1.18 \; 10^{-1}
  GeV^{3/2}; \label{res1}
\ee
\noindent  thus, the use of $B \to \pi \ell \nu_\ell$ data or 
$D \to K e \nu_e$, 
employing light and heavy flavour symmetries, give similar results.
Eqs. (\ref{res},\ref{res1}) allow to obtain, from (\ref{ratio}),
$g_{D^{*} D \gamma}$ and therefore, from (\ref{gdg}),
$\Lambda_q$ as a function of ${\hat F}$. 

We shall use below this approach as a consistency test of our method. 
In both cases, the proposed method and the consistency test, 
after having put the value of $\Lambda_q$
into (\ref{ 3}), we can compute ${\cal A}(B^* \to B \gamma)$ and the amplitude 
${\cal M}_{SD}^{(B^*)}$, i.e. the contribution of the $B^*$ pole to
the decay $B^- \to 
\mu^- {\bar \nu}_\mu \gamma$.
The final expression for the amplitude ${\cal M}_{SD}^{(B^*)}$ is as follows:
\begin{equation}
{\cal M}_{SD}^{(B^*)}= {C_1 f_{B^*} \over (v \cdot k + \Delta)} \;
\epsilon_{\mu \sigma \alpha \beta} \; \l^\mu \; \epsilon^{* \sigma} \; 
v^{\alpha} \; k^{\beta} \; , \label{5}   
\end{equation}
where $\Delta =m_{B^*}-m_B$, and $C_1$ is given by:
\begin{equation}
C_1={G_F \over \sqrt{2}} V_{ub}  {m_{B^*} \over 2 m_B } 
\sqrt{m_B m_{B^*}} \; e \; \Big[{e_b \over m_{B^*}} +
{e_q  \over \Lambda_q} \Big] \hskip 3 pt . \label{10} \end{equation}
From (\ref{5}) we can compute the contribution of the $B^*$ pole to 
$\BB(B^- \to \mu^- {\bar \nu}_\mu \gamma)$ as a function of the parameter 
${\hat F}$. 

Let us study the effect of the $B^{**}$ pole.
It is well known that in the limit of 
infinitely massive heavy quarks $(m_Q \to \infty)$ the strong dynamics of the
heavy quark decouples from the dynamics of light degrees of freedom, 
with the consequence that  the spin of the heavy quark 
and the spin of light degrees of freedom are separately conserved. 
In the case of 
the first orbitally excited heavy states, 
with orbital angular momentum 
$ \ell=1$, the total angular momentum of the light degrees of 
freedom  can be $s_\ell=1/2$ or $s_\ell=3/2$. To each of these values 
corresponds a  doublet of positive parity heavy mesons:
the doublet $(B_0,B_1^{\prime})$ with $J^P=(0^+,1^+)$
in corrispondence to $s_\ell=1/2$, and the doublet $(B_1,B_2)$ 
with $J^P=(1^+,2^+)$ ($s_\ell=3/2$).

In the charm sector only the members of the $s_\ell=3/2$ 
doublet (both in the strange and non-strange charmed channels)
have been observed: in the non-strange case they are the states 
$D_1(2420)$ and $D_2(2460)$
decaying to $D \pi$, $D^* \pi$ with pions in $D$-wave,
therefore with small decay widths  
($\Gamma_{D_1}=18^{+6}_{-4} \; MeV$ and $\Gamma_{D_2}=23\pm10 \; MeV$
\cite{PDG}). 
The states $s_\ell=1/2$ can decay into $D \pi, \; D\pi^*$
with pions in $S$-wave, and their 
large decay width makes their observation rather problematic. 
In the beauty sector, recent results from LEP Collaborations show the 
existence of positive parity orbitally excited $B$ mesons ($B^{**})$ 
with an average mass $m_{B^{**}}=5732 \pm 5 \pm 20 \; MeV$ \cite{lep}. 

Only the axial vector states $1^+$ can contribute as poles to the 
decay (Fig. 1 b); moreover, the state $B_1$, having 
$s_\ell=3/2$, has vanishing coupling to the weak current in the limit
$m_b \to \infty$ \cite{casal}, and only the state
$B^\prime_1$ (with $s_\ell=1/2$) gives a contribution in the same limit.

As to $B^\prime_1$, following the same steps leading to Eq.(\ref{5})
we get:
\begin{equation}
{\cal M}_{SD}^{(B_1^{\prime})}= i {C_2 f_{B^\prime_1} \over
(v \cdot k + \Delta^{\prime}) } 
(\epsilon \cdot v \; k_{\mu}  -v \cdot k \; \epsilon_{\mu}) 
{\it l}^{\mu}\hskip 3 pt  \label{7} \end{equation}
\noindent where $\Delta^{\prime}=m_{B^{\prime}_1}-m_B$,
\begin{equation}
<0|{\bar b} \gamma_\mu \gamma_5 q|B^\prime_1(p,\eta)>=f_{B^\prime_1}
m_{B^\prime_1}\eta_\mu \; , \label{fb1} \end{equation}
\noindent  and 
\begin{equation}
C_2={G_F \over \sqrt{2}} V_{ub} { m_{B^\prime_1}\over 2 m_B} 
\sqrt{m_B m_{B^{\prime}_1}} \; e \;
\Big[{2 e_b \tau_{1/2}(1) \over m_B} + 
{e_q  \over \Lambda_q^{\prime}} \Big] \hskip 3 pt . \label{11} 
\end{equation}
\noindent The function  $\tau_{1/2}(v \cdot v^{\prime})$ 
is the universal form factor, analogous to the IW function,
describing the matrix element of the weak current between positive 
parity heavy mesons and the doublet 
$(B,B^*)$ in the limit $m_b \to \infty$ \cite{IW1};
it can be defined as follows:
\begin{equation}
<B^\prime_1(v^\prime, \eta)|\bar b \gamma_\mu \gamma_5 b|B(v)>=\; i\; 
\tau_{1/2}(v \cdot v^\prime) \sqrt{m_B m_{B^\prime_1}}
\; \epsilon_{\mu \nu \alpha \beta}\; \eta^{* \nu} v^\alpha 
v^{\prime \beta} \hskip 3 pt , \label{tau} 
\end{equation}
\noindent 
and its value at $v \cdot v^\prime=1$ 
has been estimated by QCD sum rules \cite{paver}:
$\tau_{1/2}(1)\simeq 0.24$. We notice that, because of the
factor $1/m_B$ and the small value of 
$\tau_{1/2}(1)$, the first term in the r.h.s. of (\ref{tau}) is expected to
be small as compared to the second one.

Putting together the contributions of $B^*$ and $B_1^{\prime}$, we get:
\begin{eqnarray}
\Gamma(B^- \to \mu^- {\bar \nu}_\mu \gamma) & = &
{2 \over 3 (2\pi)^3}  \int_0^{m_B/2} dE_\gamma \;
E_\gamma^3 (m_B-2E_\gamma) 
\Bigg[ {|C_1|^2 f_{B^*}^2 \over (E_\gamma+\Delta)^2 } +
{|C_2|^2  f_{B^{\prime}_1}^2 \over 
(E_\gamma+  \Delta^{\prime})^2 } \Bigg]
\hskip 3 pt \nonumber \\
& = & \Gamma^{(B^*)} + \Gamma^{(B^\prime_1)} \; .
\label{whidth} \end{eqnarray}
\noindent

\noindent
The relative contribution of
the $B^*$ and of the $B^\prime_1$ poles is given by:
\be
{\Gamma(B^\prime_1) \over \Gamma(B^*)}={PS^{(B^\prime_1)} \over PS^{(B^*)}}
\times {|C_2 f_{B^\prime_1}|^2 \over |C_1 f_{B^*}|^2} \label{rat} \ee
\noindent therefore it is weighted by the ratio of 
the phase-space coefficients
\be
{PS^{(B^\prime_1)} \over PS^{(B^*)}} \simeq 0.54 
\ee
\noindent (using the experimental values of the mass differences
$\Delta=46 \; MeV$ and $\Delta'\simeq 500 \; MeV$) and depends on
 the ratio
$ (C_2 f_{B^\prime_1}) /( C_1 f_{B^*})$.
This last ratio is basically
determined by the coupling of the electromagnetic current to the light quarks. 
This can be seen  
from Eqs. (\ref{10},\ref{11}) neglecting $1/m_b$ terms,
since the heavy quark magnetic momentum is subleading in the inverse heavy
quark mass expansion:
\be
{C_2 \over C_1} \;
{f_{B^\prime_1} \over f_{B^*}} \simeq 
\big({m_{B^\prime_1} \over m_{B^*}}\big)^{3 \over 2} \;
{f_{B^\prime_1} \over f_{B^*}}\; {\Lambda_q \over \Lambda^{\prime}_q}  \; . 
\ee
The ratio 
${f_{B^\prime_1} \over f_{B^*}}\; {\Lambda_q \over \Lambda^{\prime}_q}$ 
can be estimated 
using experimental results from semileptonic $D-$meson decay channels
\cite{casal,noirad}. As a matter of fact, 
one can  make use of the Vector Meson Dominance (VMD) \cite{noirad} ansatz and 
assume that the coupling of the electromagnetic current to the light quarks is 
mediated by vector meson states
 $V$ ($V=\rho, \omega $).  By this assumption
the matrix elements $<B| {\bar q} \gamma^\mu q |B^*>$ 
and
$<B| {\bar q} \gamma^\mu q |B_1^{\prime}>$ can be written as:
\begin{eqnarray} 
<B(p^\prime)|{\bar q} \gamma_\mu q|P(p,\epsilon_2)>=\sum_{V,\eta}
&&<B(p^\prime)V(q,\epsilon_1(\eta))|P(p,\epsilon_2)>{i \over q^2-m_V^2}
\nonumber \\
&&<0|{\bar q} \gamma_\mu q|V(q,\epsilon_1(\eta))> \; ,\label{vmd} 
\end{eqnarray}
\noindent where $P=B^*,B^\prime_1$ and therefore they 
are proportional 
to the strong 
couplings $<B \; V |B^*>$ and $<B \; V |B_1^{\prime}>$;  these
matrix elements
have been estimated in \cite{casal} in the framework of an effective chiral 
theory for heavy mesons, using experimental information from the $D \to K^* 
\ell \nu$ semileptonic decay and the (approximate) symmetries of the effective 
theory (heavy flavour symmetry and flavour $SU(3)$ symmetry for the light 
degrees of freedom):
\be
<B(v^\prime)V(k,\epsilon)|B^*(v,\eta)>=\epsilon^{\mu \nu \alpha \beta} 
\epsilon^*_\mu \eta_\nu v_\alpha v^\prime_\beta \sqrt{m_B m_{B^*}} m_{B^*} 2 
\sqrt{2} g_V \lambda
\label{bbv} \ee
\be
<B(v^\prime)V(k,\epsilon)|B_1^\prime(v,\eta)>=[(k\cdot v)(\epsilon^* \cdot 
\eta)-(k \cdot \eta)(\epsilon^* \cdot v)] \sqrt{m_B m_{B^\prime_1}} 2 \sqrt{2} 
g_V \mu \label{bbv1} \ee
\noindent where $\lambda=-0.4\; GeV^{-1}$, $\mu=-0.13\; GeV^{-1}$ and $g_V=5.8$ 
\cite{casal}.
As a result one finds
\be
{f_{B^\prime_1} \over f_{B^*}}\; {\Lambda_q \over \Lambda^{\prime}_q}
\simeq 0.4 \;\;\; .
\ee
Therefore, we conclude that
\be
{\Gamma^{(B^\prime_1)} \over \Gamma^{(B^*)}} \simeq 0.1
\ee
\noindent i.e. the $B^\prime_1$ pole represents a contribution at the level of 
$10 \%$ of the radiative $B^- \to  \mu^- {\bar \nu}_\mu \gamma$ width.
The phase-space suppression is even larger for the contribution of 
other  (radially or orbitally) excited $B$ states, which 
therefore can be safely neglected.
Then, we can assume that the diagram of Fig. 1b adds a term of
the order of $10\%$ as compared to Fig. 1a; therefore we shall take into 
account it by multiplying the contribution of the $B^*$ pole to the
width by a normalization factor $K=1.1$. We are aware that this estimate
of the diagrams of Fig. 1b is more uncertain than the evaluation of
Fig. 1a. Nevertheless we observe that
the
VMD hypothesis successfully describes a number of low energy 
phenomena involving photon radiation \cite{marshak}; moreover these
theoretical uncertainties will not affect strongly
the final determination of $\BB(B \to \mu \nu \gamma)$, since, in any case,
the contribution of Fig. 1b is much smaller than the $B^*$ term.

The main difference of the above analysis 
with respect to ref. \cite{burdman} consists in the 
evaluation of the coupling $g_{BB^\prime_1\gamma}$ and in the role of 
$g_{B^*B\gamma}$. In \cite{burdman} 
the non relativistic quark model has been employed to compute 
$g_{B^*B \gamma}$, using 
$\Lambda_q=330 \; MeV \simeq m_u$ (constituent mass of the light quarks);
moreover, it has also been adopted to estimate 
$g_{B^\prime_1B \gamma}={g_{B^*B \gamma} \over 
\sqrt{3}}$. 
On the other hand, in our approach $g_{B^*B \gamma}$ is a 
quantity that should be inferred from experimental data, thus reducing the 
dependence of the results on the hadronization model. 
The prediction in \cite{burdman} for 
$\BB (B \to \mu \nu \gamma)$ is in the range $10^{-7}-10^{-6}$. 

Let us now come to the numerical results. 
As we observed above,
to set definite predictions one needs the experimental input 
$\Gamma(D^{*0} \to D^0 \gamma)$, which is not available yet.
In order to test the sensitivity of the method, we have used two theoretical 
determinations of the radiative $D^*$ width:
$\Gamma(D^{*0} \to D^0 \gamma)=22 \; KeV$ (ref.\cite{fdf}) and
$\Gamma(D^{*0} \to D^0 \gamma)=11 \; KeV$ (ref.\cite{cho}); with these input
data, which represent rather extreme cases (an intermediate prediction 
can be found in \cite{noirad}),
and using $V_{ub}=0.003$, 
we get the curves $(a)$ and $(b)$ depicted in Fig.2. 
From this figure  one 
can see that the branching ratio of the radiative decay mode
$B^- \to \mu^- \nu \gamma$ is expected to be larger than in the purely 
leptonic mode. This can be appreciated by comparing the prediction in Eq.(7)
with the outcome in Fig.2 in correspondence to the value 
${\hat F} \simeq 0.45 \; GeV^{3/2}$:
 $\BB(B^- \to \mu^- \nu \gamma)= 4 \; - \;  10 \; \times 10^{-7}$ 
(depending on the 
$D^{*0} \to D^0 \gamma$ decay width). This implies that the radiative mode,
for quite large values of the leptonic constant $\hat F$ is 
favoured by a factor $2 - 3$  (at least) 
with respect to the purely leptonic mode
 $B^- \to \mu^- \nu$;
moreover, as far as the statistics is concerned, 
the use of radiative decay allows to employ the 
electron channel as well, which provides a gain of an additional factor of two.

To test this result using the semileptonic data (Eqs. {\ref{res}, \ref{res1}), 
we have reported in Fig.3 two curves corresponding to the 
central values for $B \to \pi \ell \nu$ given in Eq.(\ref{bsw}) and 
(\ref{isgw});  for the range of ${\hat F}$ in Eq.(\ref{neubert}) 
we obtain 
 $\BB(B^- \to \mu^- \nu \gamma) \simeq  2 \; - \; 3 \;\times 10^{-7}$.
The data for $D \to K \ell \nu$ give an intermediate result.
As we have stressed above, the outcome in Fig.3
is based not only on experimental data, but also 
on additional theoretical assumptions, such as the polar dependence of 
the semileptonic $B \to \pi$ form factor. So, such results are characterized by 
a quite large uncertainty; nevertheless,  for 
intermediate values of $\hat F$ there is an overlap region with the result
in Fig.2 that allows us to conclude that the consistency
test does not contradict our main results in Fig.2

In conclusion, our analysis confirms that the decay channel
$B^- \to \mu^- {\bar \nu}_\mu \gamma$ can be used as a way 
to measure the 
leptonic  $B^*$ decay constant; one expects larger decay rates than in
the purely leptonic case, even though the detection of the photon 
in the final state may reduce the reconstruction efficiency. Moreover, since 
the method described here is strongly based  on HQET symmetries, 
it would be useful 
in any case to look for the radiative leptonic $B$ decay to test experimentally 
HQET predictions in this context.

\vspace*{1cm}
\par\noindent
{\bf Note added}
\par\noindent
After completing this work we became aware of the paper \cite{eil2}
where the calculation of $B \to \ell \nu \gamma$ is performed by light-cone sum 
rules. The result confirms the expectation that the radiative decay rates are 
larger than the purely leptonic rates.

\vspace*{1cm}
\par\noindent
{\bf Acknowledgments}
\par\noindent
We thank N.Paver for useful discussions.

\newpage
\begin{center} 
  \begin{Large}
  \begin{bf}
  Figure Captions
  \end{bf}
  \end{Large}
\end{center}
  \vspace{5mm}

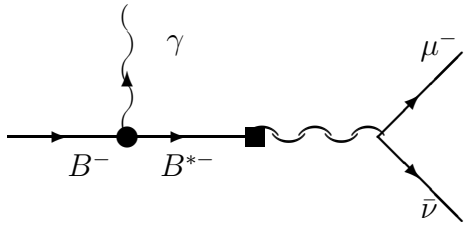
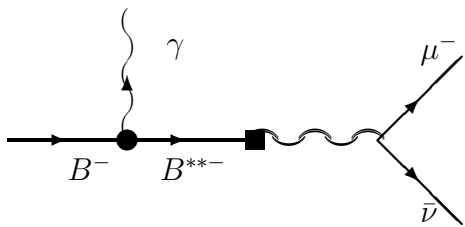
\noindent {\bf Figure 1}\\
\noindent
Diagrams dominating the $B^- \to \ell^- {\bar \nu}_\ell \gamma$ decay mode in the 
limit $m_\ell \to 0$. 
$B^*$ is the vector meson state, $B^{**}$ is the $1^+$ axial 
vector meson state.\par
\vspace{5mm}
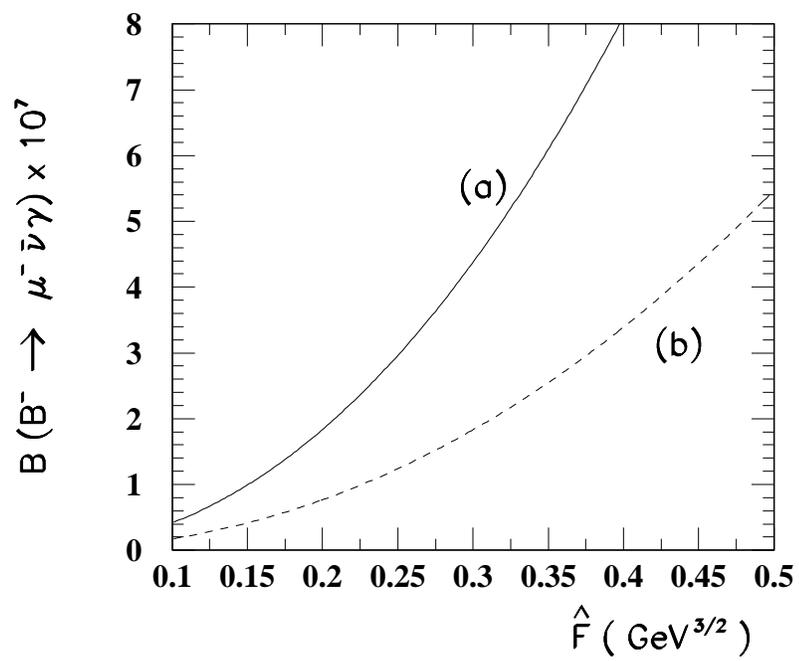
\noindent {\bf Figure 2}\\
\noindent
Branching ratio
$\BB(B^- \to \mu^- {\bar \nu}_\mu \gamma)$ obtained 
according to  the  method described in the text. 
The curves $(a)$ and $(b)$ refer to the values: 
$\Gamma(D^{*0} \to D^0 \gamma)=22 \; KeV$ \cite{fdf} (continuous line)
and $\Gamma(D^{*0} \to D^0 \gamma)=11 \; KeV$ \cite{cho} (dashed line).\par
\vspace{5mm}
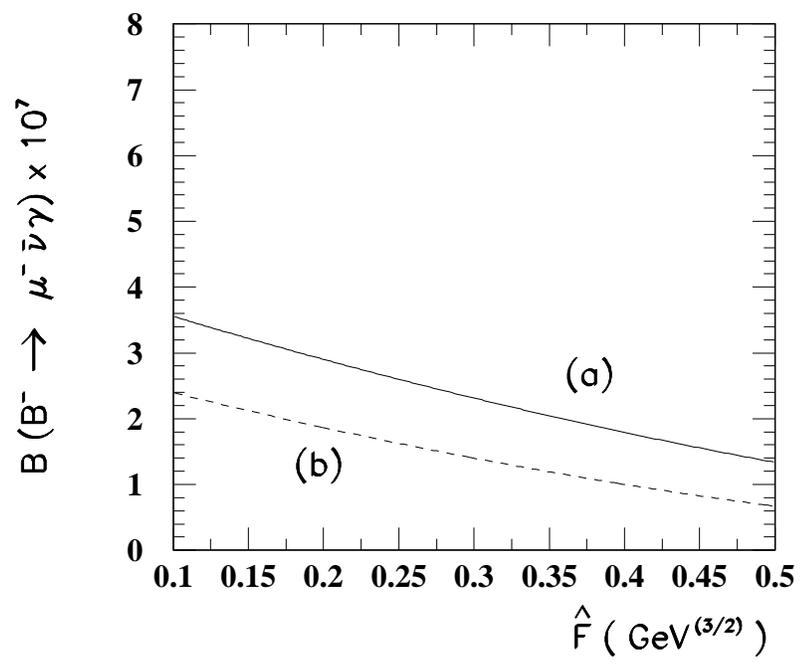
\noindent {\bf Figure 3}\\
\noindent
Branching ratio
$\BB(B^- \to \mu^- {\bar \nu}_\mu \gamma)$ obtained using data on the 
semileptonic $B \to \pi \ell \nu $ decay.
The curves $(a)$ and $(b)$ refer to the input values: 
$\BB(B \to \pi \ell \nu)=1.70 \; 10^{-4}$ (continuous line) 
and $\BB(B \to \pi \ell \nu)=1.19 \; 10^{-4}$ (dashed line).

\newpage
\begin{center}
\input FEYNMAN
\begin{picture}(24000,15000)
\THICKLINES
\drawline\fermion[\E\REG](0,0)[4500]
\drawarrow[\LDIR\ATTIP](\pmidx,\pmidy)
\global\advance\pmidy by -1500
\put(\pmidx,\pmidy){$B^-$}
\put(4500,0){\circle*{800}}
\drawline\photon[\N\REG](\pbackx,\pbacky)[5]
\drawarrow[\LDIR\ATTIP](\pmidx,\pmidy)
\global\advance\pmidx by 1500
\global\advance\pmidy by 800
\put(\pmidx,\pmidy){$\gamma$}
\drawline\fermion[\E\REG](\pfrontx,\pfronty)[4500]
\drawarrow[\LDIR\ATTIP](\pmidx,\pmidy)
\global\advance\pmidx by -1000
\global\advance\pmidy by -1500
\put(\pmidx,\pmidy){$B^{*-}$}
\put(\pbackx,\pbacky){\rule[-1.25mm]{2.5mm}{2.5mm}}
\drawline\photon[\E\REG](\pbackx,\pbacky)[5]
%\drawarrow[\LDIR\ATTIP](\pmidx,\pmidy)
\drawline\fermion[\NE\REG](\photonbackx,\photonbacky)[4500]
\drawarrow[\LDIR\ATTIP](\pmidx,\pmidy)
\global\advance\pmidy by 1500
\put(\pmidx,\pmidy){$\mu^-$}
\drawline\fermion[\SE\REG](\photonbackx,\photonbacky)[4500]
\drawarrow[\LDIR\ATTIP](\pmidx,\pmidy)
\global\advance\pmidy by -1500
\put(\pmidx,\pmidy){${\bar \nu}$}
\end{picture}
\vskip 1.5 cm
{\bf Fig. 1 a)}
\vskip 3 cm

\begin{picture}(24000,15000)
\THICKLINES
\drawline\fermion[\E\REG](0,0)[4500]
\drawarrow[\LDIR\ATTIP](\pmidx,\pmidy)
\global\advance\pmidy by -1500
\put(\pmidx,\pmidy){$B^-$}
\put(4500,0){\circle*{800}}
\drawline\photon[\N\REG](\pbackx,\pbacky)[5]
\drawarrow[\LDIR\ATTIP](\pmidx,\pmidy)
\global\advance\pmidx by 1500
\global\advance\pmidy by 800
\put(\pmidx,\pmidy){$\gamma$}
\drawline\fermion[\E\REG](\pfrontx,\pfronty)[4500]
\drawarrow[\LDIR\ATTIP](\pmidx,\pmidy)
\global\advance\pmidx by -1000
\global\advance\pmidy by -1500
\put(\pmidx,\pmidy){$B^{** -}$}
\put(\pbackx,\pbacky){\rule[-1.25mm]{2.5mm}{2.5mm}}
\drawline\photon[\E\REG](\pbackx,\pbacky)[5]
%\drawarrow[\LDIR\ATTIP](\pmidx,\pmidy)
\drawline\fermion[\NE\REG](\photonbackx,\photonbacky)[4500]
\drawarrow[\LDIR\ATTIP](\pmidx,\pmidy)
\global\advance\pmidy by 1500
\put(\pmidx,\pmidy){$\mu^-$}
\drawline\fermion[\SE\REG](\photonbackx,\photonbacky)[4500]
\drawarrow[\LDIR\ATTIP](\pmidx,\pmidy)
\global\advance\pmidy by -1500
\put(\pmidx,\pmidy){${\bar \nu}$}
\end{picture}
\vskip 1.5 cm
{\bf Fig. 1 b)}
\end{center}

\end{document}